\documentclass[preprint,showpacs,preprintnumbers,amsmath,amssymb]{revtex4}
\usepackage{booktabs}
\usepackage{mathrsfs}
\usepackage{epsfig}
\usepackage{graphicx}% Include figure files
\usepackage{dcolumn}% Align table columns on decimal point
\usepackage{bm}% bold math
\usepackage{amsmath}

\let\jnfont=\rm
\def\NPB#1,{{\jnfont Nucl.\ Phys.\ B }{\bf #1},}
\def\PLB#1,{{\jnfont Phys.\ Lett.\ B }{\bf #1},}
\def\EPJC#1,{{\jnfont Eur.\ Phys.\ Jour.\ C }{\bf #1},}
\def\PRD#1,{{\jnfont Phys.\ Rev.\ D }{\bf #1},}
\def\PRL#1,{{\jnfont Phys.\ Rev.\ Lett.\ }{\bf #1},}
\def\MPLA#1,{{\jnfont Mod.\ Phys.\ Lett.\ A }{\bf #1},}
\def\JPG#1,{{\jnfont J.\ Phys.\ G}{\bf #1},}
\def\CTP#1,{{\jnfont Commun.\ Theor.\ Phys.\ }{\bf #1},}
\def\ZPC#1,{{\jnfont Z.\ Phys.\ C }{\bf #1},}
\def\JHEP#1,{{\jnfont JHEP \ }{\bf #1},}
\def\Rv{\not{\hbox{\kern-1pt $R$}}}
\def\p{\not{\hbox{\kern-3pt $p$}}}
\begin{document}
\preprint{\parbox{1.2in}{\noindent arXive:0812.????}}

\title{\ \\[10mm]
        R-parity violating effects in top quark FCNC productions at LHC}

\author{\ \\[2mm]   Junjie Cao$^{1,2}$, Zhaoxia Heng$^3$,  Lei Wu$^1$, Jin Min Yang$^3$ }

\affiliation{
$^1$ College of Physics and Information Engineering, Henan Normal University, Xinxiang 453007, China \\
$^2$ Ottawa-Carleton Institute for Physics, Carleton University, Ottawa, K1S 5B6 Canada\\
$^3$ Institute of Theoretical Physics and Kavli Institute for Theoretical Physics China,
     Academia Sinica, Beijing 100190, China
     \vspace*{1.5cm}}

\begin{abstract}
In the minimal supersymmetric model the R-parity violating top quark interactions, which
are so far weakly constrained, can induce various flavor-changing neutral-current (FCNC)
productions for the top quark at the Large Hadron Collider (LHC).
In this work we assume the presence of the B-violating couplings and examine
their contributions to the
FCNC productions proceeding through the parton
processes $cg \to t$, $gg \to t\bar{c}$, $cg \to t\gamma$, $cg \to t Z$ and $cg \to t h$.
We find that all these processes can be greatly enhanced relative to the R-parity preserving
predictions.  In the parameter space allowed by current experiments, all the production
channels except $cg \to t h$ can reach the $3\sigma$ sensitivity, in contrast to the
R-parity preserving case in which only $cg \to t$ can reach the $3\sigma$ sensitivity.
\end{abstract}
\maketitle

\section{INTRODUCTION}
Top quark is speculated to be a sensitive probe for new physics beyond the Standard
Model (SM) since it is the heaviest fermion with a mass at weak scale  \cite{top-review}.
So far its properties are measured with a rough precision due to the limited statistics
at the Fermilab Tevatron and hence
there remains a plenty of room for new physics in top quark sector.
As a top quark factory, the LHC will scrutinize the top quark
nature, either unravelling or further constraining the new physics related
to the top quark.

Concerning the probe of new physics through the top quark, the FCNC
processes at the LHC may play an important role, just like the FCNC
transitions of the bottom quark in B-factories. The extremely
suppressed  FCNC interactions for the top quark in the SM
\cite{tcvh-sm} imply that any observation of such processes can
serve as a smoking gun for new physics. Actually, it is found that
the top quark FCNC interactions can be greatly enhanced by the
enriched flavor structure in many new physics models
\cite{top-fcnc-review}, such as the minimal supersymmetric model
(MSSM)
\cite{cao-mssm,t-decay-fcnc-mssm,t-decay-fcnc-RV,pptc-mssm,pptc-RV},
the technicolor models \cite{fcnc-tc2,tc-tc2} and the other
miscellaneous models \cite{2HDM}. Although all these models can
allow for great enhancement for the top quark FCNC processes
relative to the SM predictions, they exhibit different features and
predict very different rates for such FCNC processes. Take the
popular MSSM as an example. In the R-conserving scenario, with the
consideration of various current experimental constraints on the
parameter space, only $t \to c h $ among the FCNC decays and  $c g
\to t $ among the FCNC productions can possibly reach the observable
level at the LHC\cite{cao-mssm}. But in the presence of R-violating
couplings, top quark FCNC processes such as $t \to c V$ ($V=g, Z
,\gamma$) and $t \to c h$ can be significantly enhanced relative to
the R-conserving scenario \cite{t-decay-fcnc-RV}.

We note that while for the R-conserving MSSM  various FCNC productions
of the top quark  at the LHC  have been comprehensively studied \cite{cao-mssm},
for the  R-violating case only one FCNC production channel, i.e., $gg\to t\bar c$,
has been studied in the literature \cite{pptc-RV}. Given the central role of
the LHC in high energy physics and also considering the fact that the R-violating
top quark interactions are so far weakly constrained, it is necessary to complete
the study by examining all possible FCNC production channels induced by  R-violating
couplings. In this work we
perform such a collective study considering the productions proceeding through
the parton processes $cg \to t$, $gg \to t\bar{c}$, $cg \to t\gamma$, $cg \to t Z$
and $cg \to t h$.  We find that all these productions can be greatly enhanced
relative to the R-conserving predictions. In the parameter space allowed by
current experiments, all the production channels except $cg \to t h$ can reach
the $3\sigma$ sensitivity, in contrast to the R-parity preserving case
in which only $cg \to t$ can reach the $3\sigma$ sensitivity.

This paper is arranged as follows. In Sec. II, we recapitulate the
R-violating couplings and perform the calculations for the top
FCNC productions at the LHC. In Sec. III,  we present some
numerical results for the rates of the productions and for each
channel we show the parameter space accessible at $3\sigma$ at the
LHC. Finally,  a summary is given in Sec. IV.

\section{Calculations}
The R-violating trilinear terms in the superpotential of the MSSM,
consistent with the gauge symmetries of the SM, supersymmetry and
renormalizability, are given by \cite{rpv}
\begin{eqnarray}\label{poten}
\frac{1}{2}\lambda_{ijk}L_iL_jE_k^c +\lambda_{ijk}'L_iQ_jD_k^c
               +\frac{1}{2}\lambda^{\prime\prime}_{ijk}U_{i}^cD_{j}^cD_{k}^c
\end{eqnarray}
where $L_i(Q_i)$ and $E_i^c(U_i^c,D_i^c)$ are the left-handed
lepton (quark) doublet and right-handed lepton (quark) singlet chiral
superfields, and $i,j,k$ are generation indices.
The $\lambda_{ijk}$ and $\lambda^{\prime}_{ijk}$ violate lepton number
while $\lambda^{\prime\prime}_{ijk}$ violate baryon number.
Although it is theoretically possible to have both B-violating
and L-violating interactions, the non-observation of proton decay prohibits
their simultaneous presence. The couplings $\lambda^{\prime}$ and
$\lambda^{\prime\prime}$ can contribute to various top quark processes at the LHC:
\begin{itemize}
\item[(i)] Through exchanging a squark or slepton at tree-level, they
 can contribute
to top pair production  $pp\to t\bar t +X$  \cite{rv-tt}
and single top production $pp\to t\bar b +X$ \cite{rv-t-prod},
and also can cause some exotic top quark decays \cite{rv-t-decay}.
\item[(ii)] At loop-level they can induce top-charm FCNC interactions and
thus lead to top quark FCNC productions proceeding through the parton
processes
\begin{eqnarray}
g g \to t \bar{c}, ~c g \to t, ~c g \to t Z, ~c g \to t \gamma,
~c g \to t h \label{processes}
\end{eqnarray}
As mentioned in the preceding section, among these production
channels induced by the R-violating couplings, only  $g g \to t
\bar{c}$ has been studied in the literature. In the following we
examine these productions collectively.
\end{itemize}
Note that both the B-violating couplings $\lambda^{\prime\prime}$ and
the L-violating couplings $\lambda^{\prime}$ can induce the top quark
FCNC productions in Eq.(\ref{processes}). In our study we assume the
existence of the B-violating couplings  $\lambda^{\prime\prime}$ to show the results.
The results for the  L-violating couplings $\lambda^{\prime}$ take the similar
form and have the same behavior (e.g. the same dependence on the couplings and
the sparticle mass).

In terms of the four-component Dirac notation, the Lagrangian of the
B-violating couplings is  given by
\begin{eqnarray}
{\cal L}_{\lambda^{\prime\prime}}&=&-\frac{1}{2}\lambda^{\prime\prime}_{ijk}
\left [\tilde d^{k*}_R \bar u^{i}_R d^{jc}_L+\tilde d^{j*}_R\bar u^i_Rd^{kc}_L
       +\tilde u^{i*}_R\bar d^j_R d^{kc}_L\right ]+h.c.
\end{eqnarray}
The current upper limits on all the R-violating couplings are
summarized in \cite{R-P_v1,R-P_v}. Table I is a list of current
limits for the B-violating couplings taken from \cite{R-P_v},
which are from the analysis of $n-\bar{n}$ oscillation
\cite{lambda112}, the perturbativity requirement
\cite{perturbative} and the $Z$-decays \cite{Z-decay}. The
couplings involved in our study are $\lambda^{\prime\prime}_{2jk}$
and $\lambda^{\prime\prime}_{3jk}$, which are so far weakly
constrained. Note that these bounds are obtained for the sparticle
mass of 100 GeV and for heavier sparticles they become weak. For
example, the dependence of the $\lambda^{\prime\prime}_{323}$
bound on squark mass is 0.96 $\times$ ($m_{\tilde d_R}/100$ GeV).

%%%Table 1 %%%%%%%%%%%%%%%%%%%
\begin{table}
\caption{Current upper limits on the B-violating couplings $\lambda^{\prime\prime}_{ijk}$
         taken from \cite{R-P_v}.}
 \begin{tabular}{lll} \hline
 couplings & bounds & ~~~~~~~~~~~~~~~~sources \\
 \hline
$\lambda^{\prime\prime}_{112}$,~$\lambda^{\prime\prime}_{113}$
      ~~~~~& $10^{-6}$ & ~~~~~~~~~~~~~~~~$n-\bar{n}$ oscillation\\
 $\lambda^{\prime\prime}_{123}$,
~$\lambda^{\prime\prime}_{212}$,~$\lambda^{\prime\prime}_{213}$,~$\lambda^{\prime\prime}_{223}$
                                ~~~~~~~~~~~~& 1.25 & ~~~~~~~~~~~~~~~~perturbativity \\
$\lambda^{\prime\prime}_{312}$,~$\lambda^{\prime\prime}_{313}$
      ~~~~~& $10^{-3}$ & ~~~~~~~~~~~~~~~~$n-\bar{n}$ oscillation\\
$\lambda^{\prime\prime}_{323}$ & 0.96 & ~~~~~~~~~~~~~~~~$Z$-decays \\
\hline
 \end{tabular}\label{dlambda}
 \end{table}
%%%%%%%%%%%%%%%%%%%%%%%%%%%%%%%%%%%%%%%%%%%%%%%%%%%%%%%%%%%%%%%

In our calculations for the loop processes in Eq.(\ref{processes}),
some loop-induced vertices (like vertex $tcg$ ) appear repeatedly
in different diagrams. To simplify the calculations we define the
so-called effective vertex \cite{cao-mssm}.
For example, we define the effective $tcg$ vertex as
\begin{eqnarray}
\Gamma^{eff}_{\mu} (p_t, p_c) &=& \Gamma_\mu^{\bar{t}cg}(p_t,p_c)
      + i \Sigma (p_t) \  \frac{i (\p_t + m_c)}{m_t^2 -m_c^2} \Gamma_\mu^{\bar{c}cg}
      + \Gamma_\mu^{\bar{t}tg} \frac{i(\p_c + m_t )}{m_c^2 - m_t^2} \ i \Sigma(p_c),
\label{eff}
\end{eqnarray}
where $\Gamma_\mu^{\bar{q}qg}(q=c,t)$ is the usual QCD vertex, and
$\Gamma_\mu^{\bar{t}cg}$, $\Sigma (p_t)$ and $\Sigma(p_c)$ are
respectively the contributions from vertex and self-energy loops
shown in Fig.~\ref{vertex}, whose expressions are shown in Appendix A.
In terms of the effective $tcg$ vertex,
the Feynman diagrams for $gg\to t\bar{c}$ are shown in
Fig.~\ref{ggtc}.
In this way the analytic amplitudes are quite compact and the Fortran codes
are also simplified. Of course, we need to calculate box diagrams in Fig.\ref{ggtc}
and their expressions are presented in Appendix A.

%%Fig.1 %%%%%%%%%%%%%%%%%%%%
\begin{figure}[htb]
\scalebox{0.9}{\epsfig{file=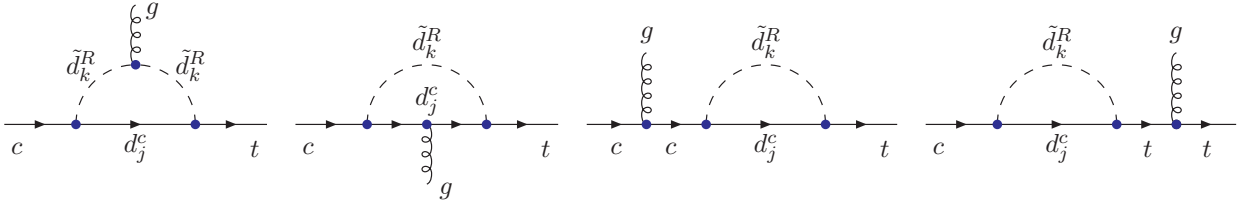}}
\vspace*{-1.5cm}
\caption{Feynman diagrams for the effective vertex $tcg$ at one-loop level.}
\label{vertex}
\end{figure}
%%%%%%%%%%%%%%%%%%%%%%%%%

%%Fig.2 %%%%%%%%%%%%%%%%%%%%
\begin{figure}[htb]
\scalebox{0.9}{\epsfig{file=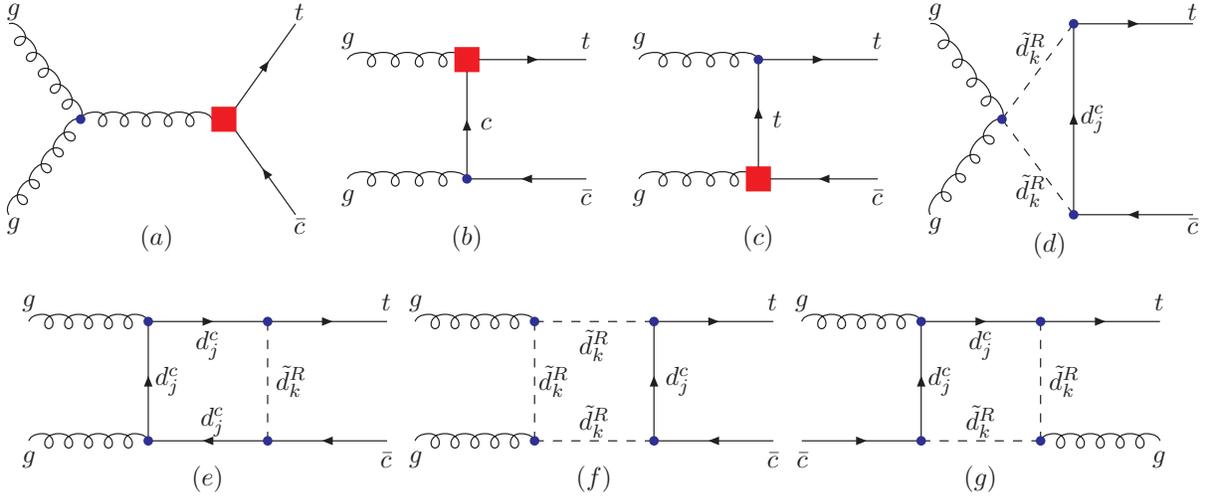}}
\vspace*{-0.7cm}
\caption{Feynman diagrams for $gg\to t\bar{c}$ at one loop level.
          The effective $tcg$ vertex in (a-c) is defined in Fig.1.}
\label{ggtc}
\end{figure}
%%%%%%%%%%%%%%%%%%%%%%%%%%%%%%

Similarly, other processes listed in Eq.(\ref{processes}) can be calculated.
Take $cg\to th$ as an example, where h is the lightest CP-even Higgs boson in the
MSSM. In order to get the effective vertex $tch$ from Fig.~\ref{vertex}, we only
need to replace the gluon with the Higgs boson in the vertex.
For the amplitude of  $cg\to th$, we can obtain it by
replacing the gluon with the Higgs boson in Fig. \ref{ggtc}
and removing the diagrams (a) and (d).
In Appendix A, we list the explicit forms for all effective vertices used
in our calculations.

\section{NUMERICAL RESULTS AND DISCUSSIONS}

The SM parameters used in our numerical calculation are \cite{pdg}
\begin{eqnarray}
m_{t}=171.2 {\rm ~GeV},~m_{Z}=91.19 {\rm ~GeV},~sin\theta_W=0.2228,
~\alpha_s(m_t)=0.1095,~\alpha=1/128.
\end{eqnarray}
The SUSY parameters involved in our calculations are the squark mass and
the B-violating couplings $\lambda^{''}_{3jk}$ and $\lambda^{''}_{2jk}$,
whose upper limits are listed in Table I.
About the constraint on squark mass, the strongest bound is from the
Tevatron experiment. For example, from the search for the inclusive
production of squarks and gluinos in R-conserving minimal
supergravity model with $A_0=0$, $\mu<0$ and $\tan\beta=5$,
the CDF gives a bound of  392 GeV at the 95 $\%$ C.L. \cite{CDF}
for degenerate gluinos and squarks. Obviously, this bound may
be not applicable to the R-violating scenario because the SUSY
signal in case of R-violation is very different from the
R-conserving case. The most robust bounds on sparticle masses come from
the LEP results, which give a bound of about 100 GeV on squark mass \cite{LEP}.
In our calculations we use CTEQ6L \cite{cteq} for parton distributions,
with the renormalization scale $\mu_R$ and factorization
scale $\mu_F$ chosen to be $\mu_R=\mu_F=m_t$.
In the following we use the parton processes to label
the corresponding hadronic processes and
all the cross sections displayed in our numerical results
are the hadronic cross sections. Also, we take into account the
charge conjugate channel for each process.

Note that our results depend on the squark mass and
the coupling product $\lambda^{''}_{3jk}\lambda^{''}_{2jk}$.
Here the product can be understood either as a single product
($\lambda^{''}_{312}\lambda^{''}_{212}$, $\lambda^{''}_{313}\lambda^{''}_{213}$ or
$\lambda^{''}_{323}\lambda^{''}_{223}$) or a sum over the indices $j$ and $k$.
In the latter case the mass degeneracy should be assumed for different down squarks
appearing in the loops in Figs.1 and 2.

%%% Fig.3 %%%%%%%%%%%%%%%%%%%%%%%%%%%%%%%%%%%%%%%%%%
\begin{figure}[htb]
\scalebox{0.67}{\epsfig{file=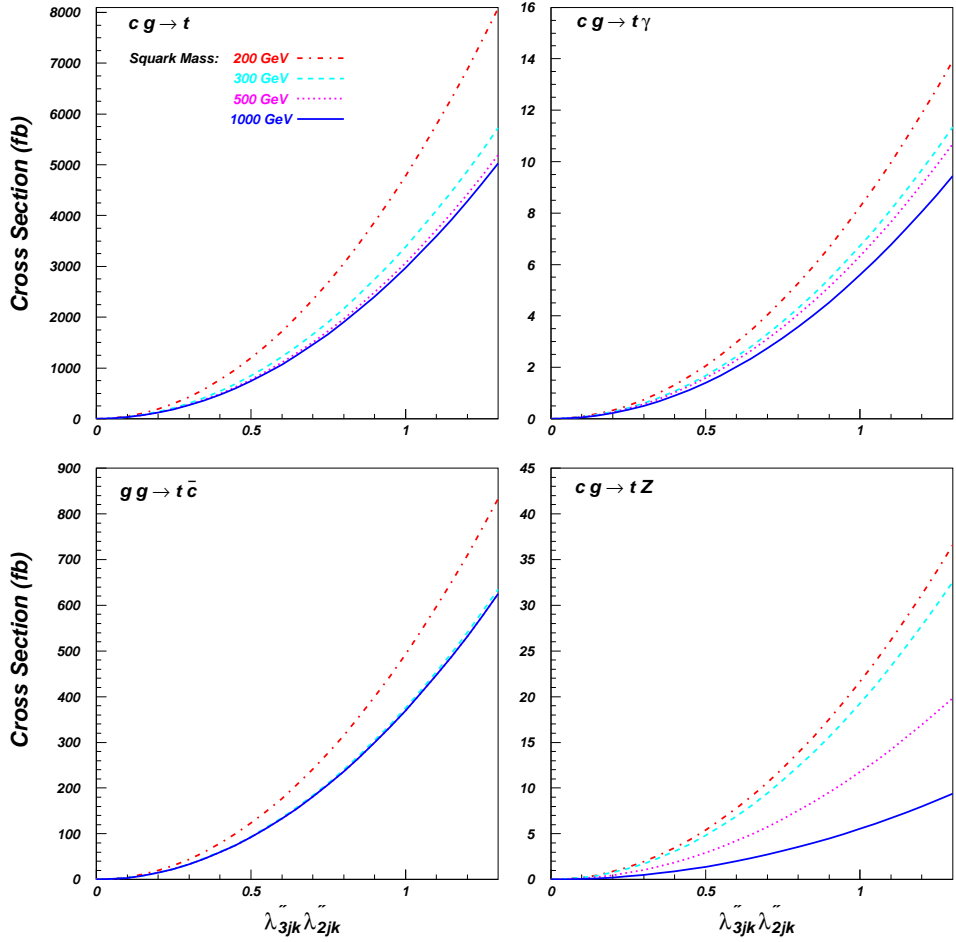}}
\vspace*{-0.8cm}
\caption{The hadronic cross sections of top quark FCNC productions at the LHC versus
         $\lambda^{''}_{3jk}\lambda^{''}_{2jk}$.}
\label{lambdafig}
\end{figure}
\begin{figure}[htb]
\scalebox{0.5}{\epsfig{file=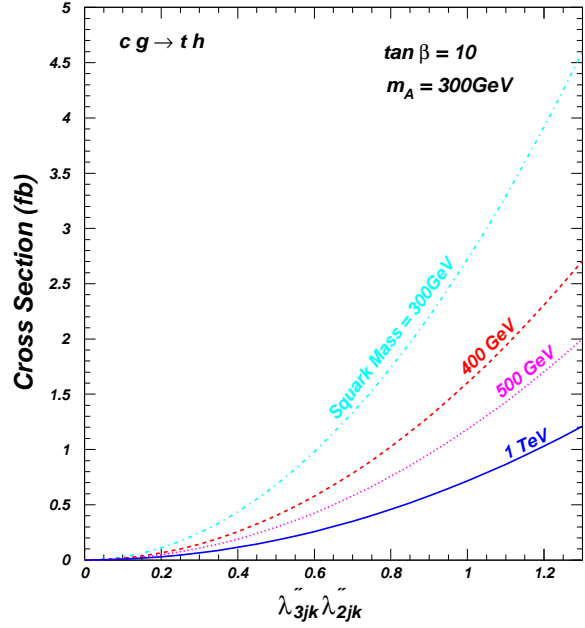}}
\vspace*{-0.5cm}
\caption{The hadronic cross section of $cg\to th$ at the LHC versus
         $\lambda^{''}_{3jk}\lambda^{''}_{2jk}$ .}
\label{cgth_lambda}
\end{figure}

%%%%%%%%%%%%%%%%%%%%%%%%%%%%%%%%%%%%%%%%%%%
In Figs.\ref{lambdafig} and \ref{cgth_lambda} we plot the hadronic cross sections
versus $\lambda^{''}_{3jk}\lambda^{''}_{2jk}$ for different squark masses.
From the figures we see that cross sections increase with
$|\lambda^{''}_{3jk}\lambda^{''}_{2jk}|^2$ and decrease with squark mass.
Note that in Fig. \ref{cgth_lambda} the mass of the Higgs boson $h$
is determined by $M_A$, $\tan\beta$ and the varying squark mass (we assume
the mass degeneracy for all squarks including the top squarks). For the
parameters chosen in Fig. \ref{cgth_lambda},  the Higgs boson $h$ is slightly
above 100 GeV.

%%%%% Fig.4 %%%%%%%%%%%%%%%%%%%%%%%%%%%%%%%%%%%%%%%%%%%%%
%%%%%%%%%%%%%%%%%%%%%%%%%%%%%%%%%%%%%%%%%%%%%%%%%%%%%%%%%%%%%%
%%%%%Table II %%%%%%%%%%%%%%%%%%%%%%%%%%%%%%%%%%%%%%%%
\begin{table}
\caption{\small
The hadronic cross sections of top quark FCNC productions
  at the LHC in R-violating MSSM  (for squark mass of 300 GeV
and the value of $\lambda^{''}_{3jk}\lambda^{''}_{2jk}$ summed
over $j,k$ with $\lambda^{''}_{3jk}=1$ and $\lambda^{''}_{2jk}=1.25$ )
in comparison with the maximal values in the R-conserving MSSM and
the top-color assisted technicolor (TC2) model.
  The corresponding charge-conjugate channels are also included.
  The LHC $3 \sigma$ sensitivities in the last column are
  estimated for an integrated luminosity of 100 fb$^{-1}$.}
\vspace*{0.2cm}
\begin{tabular}{|l|l|l|l|l|} \hline
~  & \multicolumn{2}{c|}{MSSM} & ~~TC2~~ &~~LHC $3 \sigma$ sensitivity \\ \cline{2-3}
~  &~R-conseving~&~R-violating~ &~~  & \\ \hline
~$gg \to t\bar{c}$ &~~ 700 fb \cite{cao-mssm} & ~~5 pb &~~30 pb \cite{fcnc-tc2}&~~ 1500 fb \cite{pptc-bkg}\\ \hline
~$cg \to t$        &~~ 950 fb \cite{cao-mssm} & ~~47 pb  &~~1.5 pb \cite{fcnc-tc2} &~~ 800 fb \cite{hosch} \\ \hline
~$cg \to t \gamma$ &~~ 1.8 fb \cite{cao-mssm} & ~~94 fb  &~~20 fb  \cite{fcnc-tc2} &~~ 5 fb \cite{aguila} \\ \hline
~$cg \to tZ$~      &~~ 5.7 fb \cite{cao-mssm} & ~~305 fb  &~~100 fb \cite{fcnc-tc2}&~~ 35 fb \cite{aguila}  \\ \hline
~$cg \to th$~      &~~ 24 fb \cite{cao-mssm}  & ~~37 fb   &~~600 fb \cite{fcnc-tc2}  &~~ 200 fb \cite{th-aguilar}\\ \hline
\end{tabular}
\end{table}
%%%%%%%%%%%%%%%%%%%%%%%%
In Table II we display the hadronic cross sections
in R-violating MSSM in comparison with the results in the top-color assisted
technicolor (TC2) and the R-conserving MSSM. Also we listed the 3$\sigma$ sensitivity
for each production channel at the LHC with a luminosity of 100 fb$^{-1}$.
We see that the R-violating couplings allow for much larger cross sections
than the R-conserving MSSM and all channels except $cg\to th$ can reach
the  3$\sigma$ level in the allowed parameter space.

\begin{figure}[tb]
\scalebox{0.6}{\epsfig{file=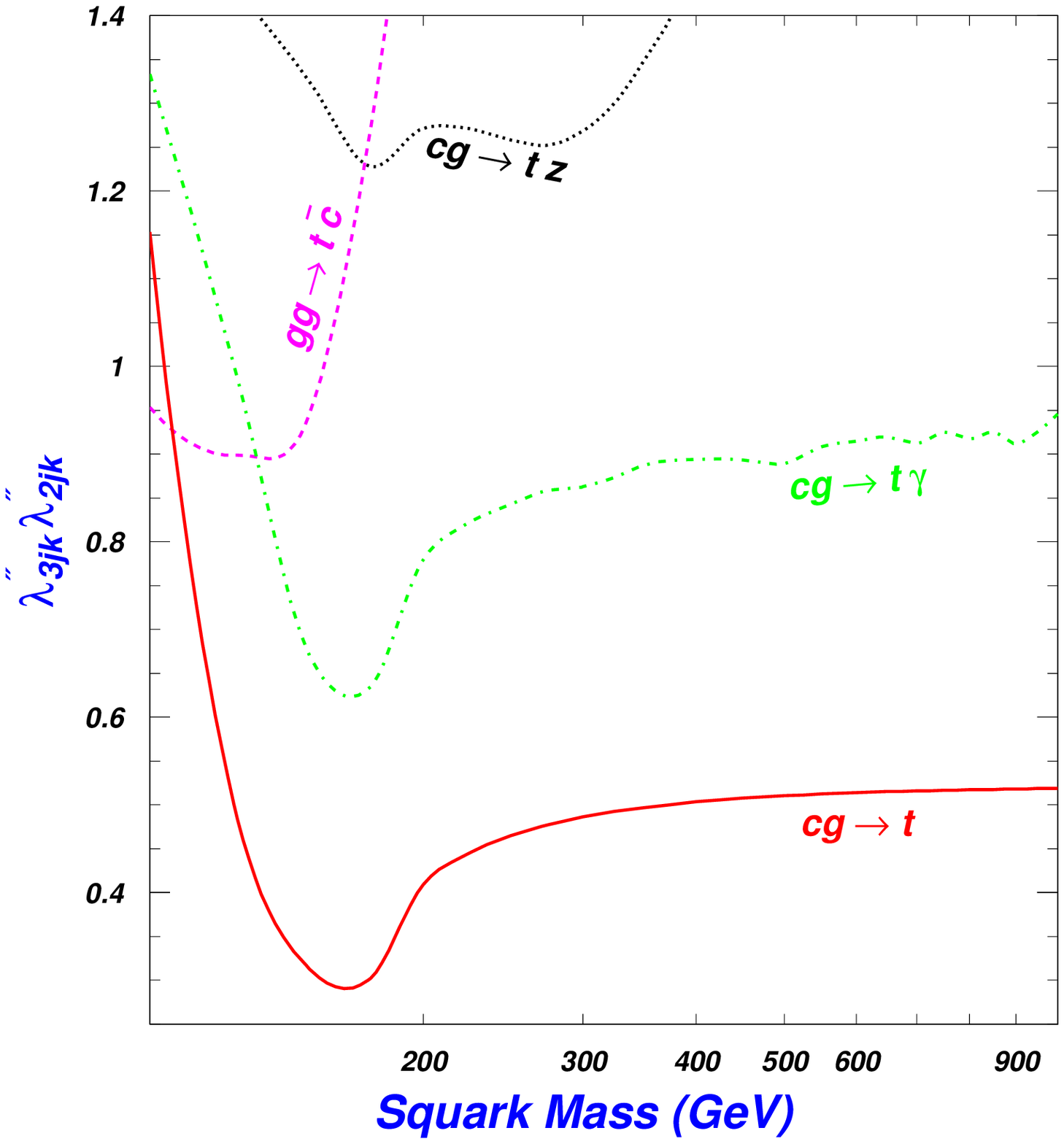}}
\vspace*{-0.4cm}
\caption{The 3$\sigma$ contour of the hadronic cross sections at the LHC.
         The region above each curve is the corresponding  3$\sigma$ observable region.}
\label{fcncfig}
\end{figure}

In Fig.\ref{fcncfig} we plot the 3$\sigma$ contours of the
hadronic cross sections in the plane of
$\lambda^{''}_{3jk}\lambda^{''}_{2jk}$ versus squark mass. The
region above each curve is the corresponding observable region at
3$\sigma$. We see that among these channels the production
proceeding through $cg\to t$ is most powerful for probing such
R-violating SUSY. The peaks of the curves near the top quark mass
show the resonance behavior of the top quark self-energy which
involve a squark and a light quark in the loops.
%%%% Fig.5 %%%%%%%%%%%%%%%%%%%%%%%%%%%%%%%%%%%%%%%%%%%%%%%
%%%%%%%%%%%%%%%%%%%%%%%%%%%%%%%%%%%%%%%%%%%%%%%%%%%%%%%%%%%%%%%%%%%%%%%%%%%%

Finally, we remark that there may be some correlation between top quark
FCNC interactions and $b$ or $c$ physics, as discussed by using the
model-independent effective operators \cite{final}.
In case of B-violating couplings, the co-existence of two couplings
can induce some $b$ or $c$ processes. For example,
the co-existence of $\lambda^{''}_{312}$ and $\lambda^{''}_{313}$
can induce $b\to s $ transition and thus the product
$\lambda^{''}_{312} \lambda^{''}_{313}$ is constrained by $b\to s \gamma$.
A complete list of such bounds on the product of two couplings
from $b$ or $c$ physics is presented in \cite{R-P_v}.
For the product $\lambda^{''}_{3jk}\lambda^{''}_{2jk}$ involved
in our study, it is not constrained by those $b$ or $c$ processes
since the co-existence of  $\lambda^{''}_{3jk}$ and $\lambda^{''}_{2jk}$
cannot trigger those processes (in other words, for $\lambda^{''}_{3jk}$
or $\lambda^{''}_{2jk}$ to trigger those processes,
another coupling different from these two must be present).

\section{CONCLUSION}
In the MSSM the R-violating top quark interactions are so far
weakly constrained, which can induce various FCNC productions for
the top quark at the LHC. We assumed the presence of the
B-violating top quark couplings and examined the induced FCNC
productions which proceed through the parton-level processes $cg
\to t$, $gg \to t\bar{c}$, $cg \to t\gamma$, $cg \to t Z$ and $cg
\to t h$. We found that all these processes can be greatly
enhanced relative to the R-parity preserving predictions.  In the
parameter space allowed by current experiments, all the production
channels except $cg \to t h$ can reach the $3\sigma$ sensitivity,
in contrast to the R-parity preserving case in which only $cg \to
t$ can reach the $3\sigma$ sensitivity. Recall that among the FCNC
decays of the top quark, only $t\to ch $ could marginally be
accessible at the LHC in the R-conserving MSSM \cite{cao-mssm},
while in the R-violating MSSM all the FCNC decay modes could reach
the observable level at the LHC \cite{t-decay-fcnc-RV}. So, if
supersymmetry is proven to the true story at the LHC, these FCNC
productions and decays of the top quark could shed some light on
R-parity conservation or violation.

\section*{Acknowledgement}
This work was supported in part by the National Natural Science
Foundation of China (NNSFC) under grant Nos. 10505007, 10821504,
10725526 and 10635030, by HASTIT under grant No. 2009HASTIT004,
and  by the National Sciences and Engineering Research Council of
Canada.

\appendix
\section{Expressions of loop results}
Here we list the expressions for $\Sigma(p)$ and $\Gamma_\mu^{\bar{t}cg}$ in the
effective $tcg$ vertex of Eq.(\ref{eff}), and also list the expressions for
$\Gamma_\mu^{\bar{t}c\gamma}$, $\Gamma_\mu^{\bar{t}cZ}$ and $\Gamma^{\bar{t}ch}$
appearing respectively in the effective $tc\gamma$,$tcZ$ and $tch$ vertex.
Their expressions are given by
\begin{eqnarray} \label{A1}
\Sigma(p)&=& aB^1_\alpha\gamma^\alpha P_R ~,\\
\Gamma_\mu^{\bar{t}cg}&=& ag_s[C^1_{\alpha\beta}\gamma^\alpha\gamma_\mu\gamma^\beta
    -C^1_\alpha\gamma^\alpha\gamma_\mu (p_t\!\!\! \!\slash +p_c\!\!\! \!\slash )
    -2C^2_{\mu\alpha}\gamma^\alpha + C^2_\alpha\gamma^\alpha (p_t - p_c)_\mu ]P_R\\
\Gamma^{\bar{t}ch}&=& -ae \left\lbrace  Y_dm_{d_j}[ 2C^3_\alpha\gamma^\alpha
     + C^3_0 (p_c\!\!\! \!\slash - p_t\!\!\! \!\slash) ]
     + Y_{\tilde{d}_R}C^4_\alpha\gamma^\alpha \right\rbrace  P_R~, \label{A3} \\
\Gamma_\mu^{\bar{t}c\gamma}&=& -\frac{2}{3}ae[C^1_{\alpha\beta}\gamma^\alpha\gamma_\mu\gamma^\beta
    -C^1_\alpha\gamma^\alpha\gamma_\mu (p_t\!\!\! \!\slash +p_c\!\!\! \!\slash )
    -2C^2_{\mu\alpha}\gamma^\alpha + C^2_\alpha\gamma^\alpha (p_t - p_c)_\mu ]P_R\\
    \Gamma_\mu^{\bar{t}cZ}&=& ae\{ (v_f+a_f) [C^5_{\alpha\beta}\gamma^\alpha\gamma_\mu\gamma^\beta
    -C^5_\alpha\gamma^\alpha\gamma_\mu (p_t\!\!\! \!\slash +p_c\!\!\! \!\slash )] \nonumber \\
&&-b[2C^6_{\mu\alpha}\gamma^\alpha - C^6_\alpha\gamma^\alpha (p_t - p_c)_\mu ] \} P_R~,
\label{A5}
\end{eqnarray}
with $p_t$ and $p_c$ denoting respectively the momenta of the top and charm quark,
and the constants given by
\begin{eqnarray}
&& a=\frac{i}{16\pi^2}\lambda^{''}_{3jk}\lambda^{''}_{2jk}~,
   ~~b=-\frac{\sin\theta_W}{3\cos\theta_W},~\\
&& a_f=-\frac{1}{4\sin\theta_W\cos\theta_W}~,
   ~~v_f=a_f(1-\frac{4}{3}\sin^2\theta_W)~.
\end{eqnarray}
For the loop functions $B$ and $C$ in Eqs.(\ref{A1}-\ref{A5}), we
adopted the definitions in \cite{loop} and use LoopTools \cite{Hahn}
in the calculations. The loop functions' dependence on the momenta
and the masses is given by
\begin{eqnarray}
C^1&=&C(-p_t-p_c,p_c,m^2_{d_j},m^2_{d_j},m^2_{\tilde{d}_k^R})~,\\
C^2&=&C(-p_t,p_t+p_c,m^2_{d_j},m^2_{\tilde{d}_k^R},m^2_{\tilde{d}_k^R})~,\\
C^3&=&C(p_c-p_t,-p_c,m^2_{d_j},m^2_{d_j},m^2_{\tilde{d}_k^R})~,\\
C^4&=&C(-p_t,p_t-p_c,m^2_{d_j},m^2_{\tilde{d}_k^R},m^2_{\tilde{d}_k^R})~,\\
C^5&=&C(-p_t,p_c,m^2_{d_j},m^2_{d_j},m^2_{\tilde{d}_k^R})~,\\
C^6&=&C(-p_t,p_t-p_c,m^2_{d_j},m^2_{\tilde{d}_k^R},m^2_{\tilde{d}_k^R})~,\\
B^1&=&B(-p,m^2_{\tilde{d}_k^R},m^2_{d_j} )~.
\end{eqnarray}
The expressions for Yukawa couplings $Y_d$ and $Y_{\tilde{d}_R}$ in
Eqs.(\ref{A3}) are given by
\begin{eqnarray}
Y_d&=& \frac{m_d\sin\alpha}{2m_W\sin\theta_W\cos\beta} ,\\
Y_{\tilde{d}_R}&=& -\frac{1}{3}m_Z\tan\theta_W\sin(\alpha+\beta)
+\frac{m^2_d\sin\alpha}{m_W\sin\theta_W\cos\beta} .
\end{eqnarray}
The amplitudes of the box diagrams in Fig.\ref{ggtc}(a-g)
are given respectively by
\begin{eqnarray}
M_{(d)}&=&-ag_s^2T_1\varepsilon^a_\rho(p_1) \varepsilon^b_\sigma(p_2)g^{\rho\sigma}
    \bar{u}(p_t)(C_\alpha\gamma^\alpha)P_Rv(p_c)~,\\
M_{(e)}&=&ag_s^2T_2\varepsilon^a_\rho(p_1) \varepsilon^b_\sigma(p_2)
    \bar{u}(p_t)[D^1_{\alpha\beta\delta}\gamma^\alpha\gamma^\rho\gamma^\beta\gamma^\sigma\gamma^\delta
    -D^1_{\alpha\beta}\gamma^\alpha\gamma^\rho p_1\!\!\! \!\slash\gamma^\sigma\gamma^\beta \nonumber\\
    &&-D^1_{\alpha\beta}\gamma^\alpha\gamma^\rho\gamma^\beta\gamma^\sigma(p_c\!\!\! \!\slash + p_t\!\!\! \!\slash)
    +D^1_\alpha\gamma^\alpha\gamma^\rho p_1\!\!\! \!\slash\gamma^\sigma(p_c\!\!\! \!\slash
        + p_t\!\!\! \!\slash)]P_Rv(p_c)~,\\
M_{(f)}&=&ag_s^2T_2\varepsilon^a_\rho(p_1) \varepsilon^b_\sigma(p_2)
    \bar{u}(p_t)[4D^2_{\rho\sigma\alpha}\gamma^\alpha
    +2D^2_{\rho\alpha}\gamma^\alpha(2p_1+p_2-2p_t)_\sigma \nonumber\\
    &&-2D^2_\sigma\alpha\gamma^\alpha(2p_t-p_1)_\rho
    -D^2_\alpha\gamma^\alpha(2p_t-p_1)_\rho(2p_c-p_2)_\sigma]P_Rv(p_c)~,\\
M_{(g)}&=&ag_s^2T_3\varepsilon^a_\rho(p_1) \varepsilon^b_\sigma(p_2)
    \bar{u}(p_t)[2D^3_{\sigma\alpha\beta}\gamma^\alpha\gamma^\rho\gamma^\beta
    -2D^3_{\sigma\alpha}\gamma^\alpha\gamma^\rho p_1\!\!\! \!\slash \nonumber\\
    &&-D^3_{\alpha\beta}\gamma^\alpha\gamma^\rho\gamma^\beta(2p_t-p_2)_\sigma
    +D^3_\alpha\gamma^\alpha\gamma^\rho p_1\!\!\! \!\slash(2p_t-p_2)_\sigma]P_Rv(p_c)~,
\end{eqnarray}
where
\begin{eqnarray}
T_1=\varepsilon_{mlk}\varepsilon_{nli}(T^aT^b+T^bT^a)_{ik}~ ,
T_2=\frac{1}{2}\delta_{ab}\delta_{mn}-(T^bT^a)_{mn}~ ,
T_3=\varepsilon_{mlk}\varepsilon_{nij}T^a_{il}T^b_{jk}~ ,
\end{eqnarray}
with $m,n,a,b$ being respectively the color indices of the top, charm
and the two gluons, and $p_1$ and $p_2$ being the momenta of the two gluons.
The loop functions' dependence is given by
\begin{eqnarray}
C&=&C(-p_t,p_t+p_c,m^2_{d_j},m^2_{\tilde{d}_k^R},m^2_{\tilde{d}_k^R}) ~,\\
D^1&=&D(-p_t,-p_c,p_2,m^2_{d_j},m^2_{\tilde{d}_k^R},m^2_{d_j},m^2_{d_j}) ~,\\
D^2&=&D(-p_t,p_1,p_2,m^2_{d_j},m^2_{\tilde{d}_k^R},m^2_{\tilde{d}_k^R},m^2_{\tilde{d}_k^R}) ~,\\
D^3&=&D(-p_t,p_2,-p_c,m^2_{d_j},m^2_{\tilde{d}_k^R},m^2_{\tilde{d}_k^R},m^2_{d_j}) ~.
\end{eqnarray}

\end{document}